\documentclass{ctextemp_elsart}

\usepackage{graphicx}
\usepackage{amssymb}

\begin{document}

\begin{frontmatter}



\title{A simple scheme for quantum networks based on orbital angular momentum states of photons\thanksref{talk}}
\thanks[talk]{Project supported by the State Key Development Program for Basic Research of China (Grant No.2007CB307001). }
\author{Zhi-Kun Su, }
\author{Fa-Qiang Wang\corauthref{cor}},
\author{ Rui-Bo Jin, Rui-Sheng Liang, Song-Hao Liu}
\corauth[cor]{Correspondent author:fq\_wang@163.com}
\address{Laboratory of Photonic Information Technology,School for Information and Optoelectronic Science and Engineering, South China Normal University,Guangzhou 510006, PR China}

\begin{abstract}
We propose a new quantum network scheme using orbital angular
momentum states of photons to route the network and spin angular
momentum states to encode the information. A four-user experimental
scheme based on this efficient quantum network is analyzed in
detail, which is particularly appealing for the free space quantum
key distribution.Users can freely exchange quantum keys with each
other.
\end{abstract}

\end{frontmatter}

\section{Introduction}
Quantum key distribution (QKD) network has recently attracted
growing attention. In the past few years, researchers have proposed
several outstanding QKD network schemes based on different
mechanisms, such as the passive optical network (PON)[1], the
silica-based planar lightwave circuit (PLC) non-blocking matrix
switch[2], the wavelength division multiplexing (WDM)[3][4], and the
multi-particle entanglement[5] etc. The passive optical network[1]
opens a way for the development of quantum network, though it's not
a practical implementation since the network users have to share
security information with a single controller. And the PLC matrix
switch network[2] shows the possibility of sending quantum signals
through current optical networks. Most recently, Wei.C and his
colleagues[4] introduce a quantum router which processes network
addresses through WDM. Anyway, the common feature of these schemes
is that they aim to build a one-to-any QKD network or an any-to-any
one through the conventional optical network. In this paper, we
present a novel network mechanism using spin angular momentum (SAM)
and orbital angular momentum (OAM) [6] states of photons, which is
particularly appealing for the free space QKD. The angular momentum
of photons can contain not only a spin contribution associated with
the polarization, but also an orbital contribution arising from
phase fronts that are inclined with respect to the beam's
propagation axis. Such screw beams have an azimuthal phase
dependence of the form $exp(i\ell\phi)$ and carry a discrete OAM of
$\ell\hbar$ units per photon along their propagation direction. The
value $\ell$ is an integer and corresponds to the number of times
the phase changes by $2\pi$  in a closed loop around the beam. In
2004, it was reported that these OAM states can be used to transmit
information [7]. Most recently, L.-P.Deng et al. have present
two-optical CNOT gates using SAM and OAM of a single photon as
qubits [8]. In our newly network scheme, the quantum information is
encoded in SAM states of photons, and the network addressing is
automatically processed according to OAM states. Users can directly
exchange the quantum keys with any other without a controller.Since
the Hilbert space spanned by OAM states is in principle infinite,
the number of users in this network can be increasable, but one
point we should note is that the number of channels depends on the
size of every aperture[9].

The paper is organized as follows: in Section 2 we review the
interferometric OAM sorters[10][11] which can distinguish the OAM
states of single photons. Section 3 analyzes in detail an
experimental network scheme based on the newly-proposed total
angular momentum (TAM) sorter[12]. Section 4 gives a summary of the
paper.
 \label{}
\section{Interferometric OAM sorter}
In this section, we will use matrix formulation[13][14] to depict
the propagation property of photons with definite OAM through the
previous interfermetric OAM sorters. Laguerre-Gaussian (LG) beams
are examples of light beams with an intensity structure that is
rotationally symmetric about the beam axis and a phase structure
with an azimual dependence $exp(i\ell\phi)$. Any LG modes of order N
can be expanded as the sum of $(N+1)$ Hermite-Gaussian (HG) modes of
the same order:
\begin{equation}
u_{lp}^{LG}=u_{nm}^{LG}=\sum\limits_{k=0}^{N}a(n,m,k)u_{N-k,k}^{HG}(x,y,z),
\end{equation}
with the complete coefficients:
\begin{equation}
a(n,m,k)=i^{k}[\frac{(N-k)!k!)}{2^{N}n!m!}]^{1/2}\frac{1}{k!}\frac{d^{k}}{dt^{k}}[(1-t)^{n}(1+t)^{m}]_{t=0},
\end{equation}
where $n$, $m$ are arbitrary nonnegative integers, and
$N=n+m=2p+|\ell|$ is the order of the mode. Laguerre-Gaussian modes
are characterized by two mode indices $l=|n-m|$ and $p=min(m,n)$,
where $\ell$ is the number of $2\pi$ cycles in phase around the
circumference and $p+1$ is the number of radial nodes. It follows
that such a mode can be represented by a column vector with $(N+1)$
elements:
\begin{equation}
|u_{lp}^{LG}\rangle=|u_{nm}^{LG}\rangle=\left(
                                          \begin{array}{c}
                                            a(n,m,0) \\
                                            a(n,m,1) \\
                                            \cdots \\
                                            \cdots \\
                                            a(n,m,N-1) \\
                                            a(n,m,N) \\
                                          \end{array}
                                        \right)
.
\end{equation}
To show how this $(N+1)$-element column vector may be used, let's
consider the passage of the LG mode of order N through a beam
rotator. The $[(N+1) ¡Á~(N+1)]$ mode rotation matrices
$[rot(\alpha)]_{N}$ of a beam rotator are given in Refs.[13][14].
Take $\ell=2$, $p=0$ for example, the mode
$|u_{l=2,p=0}^{LG}\rangle$ passes through a beam rotator with
rotation angle $\alpha$, the column vector describing the output
beam is given by:
\begin{equation}
[rot(\alpha)]_{N=2}|u_{l=2,p=0}^{LG}\rangle=\left(
                                              \begin{array}{ccc}
                                                \cos^{2}\alpha & \frac{\sin2\alpha}{\sqrt{2}} & \sin^{2}\alpha \\
                                                -\frac{\sin2\alpha}{\sqrt{2}} & \cos2\alpha & \frac{\sin2\alpha}{\sqrt{2}} \\
                                                \sin^{2}\alpha & -\frac{\sin2\alpha}{\sqrt{2}} & \cos^{2}\alpha \\
                                              \end{array}
                                            \right)
 \left(
   \begin{array}{c}
     \frac{1}{2} \\
     \frac{-i}{\sqrt{2}} \\
     \frac{-1}{2} \\
   \end{array}
 \right)
 =e^{-2i\alpha}|u_{l=2,p=0}^{LG}\rangle.
\end{equation}
As we can see from Eq.(4) , the relative phase difference between
the input and output beam is $\Delta\Psi=2\alpha$. Furthermore, when
the LG mode with the azimuthal phase form $exp(i\ell\phi)$ passes
through a rotator with angle $\alpha$, the phase shift between input
and output mode is $\Delta\Psi=\ell\alpha$. Therefore, for
particular combinations of $\ell$ and $\alpha$, the rotated beam may
be either in or out of phase with respect to the original. If such a
rotation is incorporated into the arms of a two-beam interferometer,
then the phase shift between the two arms becomes $\ell$ dependent.
Using the above concept, Leach etc. proposed a sorter for
distinguishing the OAM of single photons[10]. It consists of
cascading Mach-Zender interferometers with a Dove prism placed
appropriately in each arm. A simple diagram of the first stage of
this sorter is presented in Fig.1(a). The two Dove prisms, rotated
with respect to each other through an angle $\alpha/2$, rotate a
passing beam through an angle $\alpha$. In 2003 a simplification of
the above approach was described [11]. Comparatively, the sketch of
the latter approach is outlined in Fig.1(b). The phase difference
introduced by the beam rotator is $\ell\alpha$. The phase offset
introduced by the phase delay plate is independent of $\ell$. This
device can sort individual photons according to their $\ell$ values,
since the photons exit the interferometer through different output
ports. Ref.[11] also pointed out a quite original idea that we can
utilize the distinguishable OAM states as distinct channels. Such a
scheme may be called mode-division multiplexing (MDM), in analogy
with the traditional terminology WDM of optical communications.
\begin{figure}
  \includegraphics{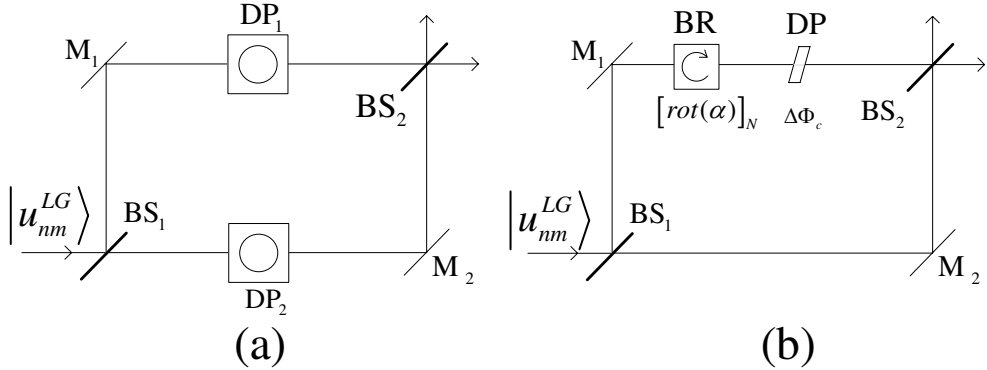}
  \centering
  \caption{First stage of an OAM sorter that separates photons according to their OAM states. (a)Two Dove prisms are placed in each arm of the Mach-Zender interferometer respectively.(b)A beam rotator (rotation angle $\alpha$) and a delay plate (phase shift $\Delta\Phi_{c}$) are inserted in one arm of the Mach-Zender interferometer.}\label{Fig.1. }
\end{figure}
\label{}
\section{OAM quantum network}
After describing the feature of OAM sorter, we will discuss how to
use OAM sorter to realize quantum network. Let's first analyze the
experimental network scheme based on the newly-proposed TAM sorter
in this section. The TAM of a photon is given by the sum of SAM and
OAM. In the network, we encode information in the SAM of photons and
use distinguishable OAM states to process network addresses.
Combining the interferometric OAM sorter[10] with proper
polarization beam splitters (PBS) provides a way to sort photons on
the basis of TAM. However, in general a Dove prism slightly changes
the polarization state of a passing beam and therefore cannot be
used for TAM sorting. For the sake of simultaneously measuring the
SAM and OAM of single photons exactly, Leach etc. designed a prism
that rotates both phase and intensity but acts as a quarter-wave
plate on the polarization vector[12]. Under this excellent
achievement, we devised a new scheme for QKD network. To illustrate
our idea more clearly, we give an example of four-user QKD as shown
in Fig.2.
\begin{figure}
  \includegraphics{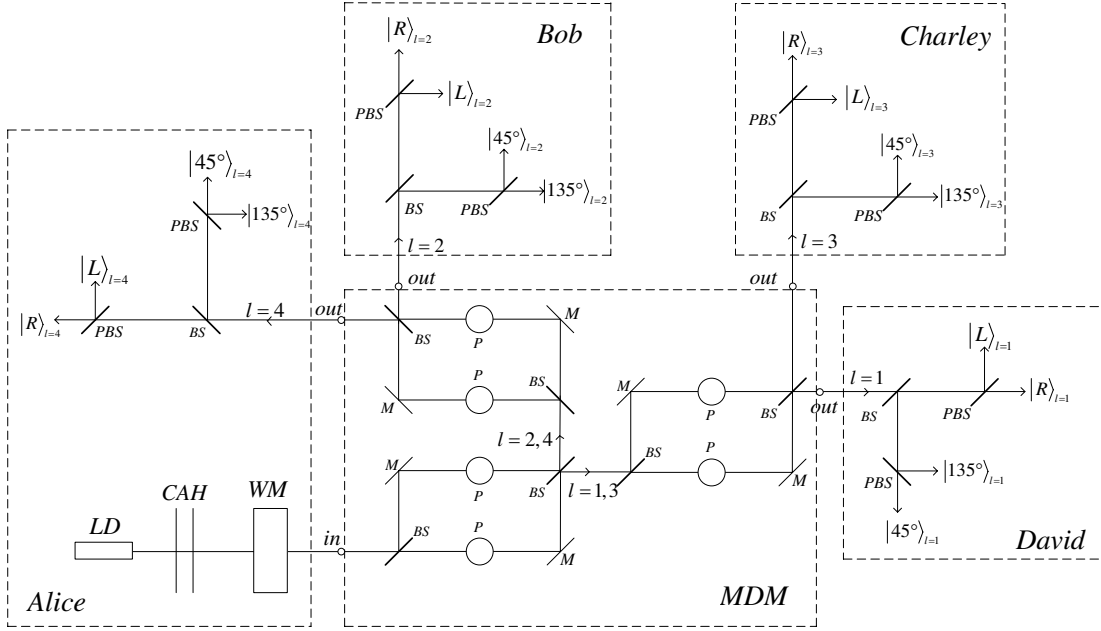}\\
  \caption{The OAM-QKD scheme: LD, laser diode; CAH, computer addressable hologram; WM, wave modulator; BS, beam splitter; P, prism designed by Leach etc.[12]; PBS, polarizing beam splitter; M, mirror.}\label{Fig.2 }
\end{figure}
From Fig.2, we can see that Alice is connected to Bob, Charley and
David by way of the MDM. When Alice wants to communication with Bob,
Charley or David, she uses the fist diffraction order of a computer
addressable hologram (CAH) to prepare photons of $\ell=2$, $\ell=3$
or $\ell=1$ accordingly. These photons will be separated by the OAM
sorters and transmitted to the appointed users according to the OAM
states of photons.

In this scheme, photons are prepared in continuum of polarization
states, for example, the two diagonal states,
$|45^\circ\rangle=\frac{1}{\sqrt{2}}\left(
                                    \begin{array}{c}
                                      1 \\
                                      1 \\
                                    \end{array}
                                  \right)$ and $|135^\circ\rangle=\frac{1}{\sqrt{2}}\left(
                                    \begin{array}{c}
                                      1 \\
                                      -1 \\
                                    \end{array}
                                  \right)$, or the two
circular states, $|L\rangle=\frac{1}{\sqrt{2}}\left(
                                    \begin{array}{c}
                                      1 \\
                                      i \\
                                    \end{array}
                                  \right)$ and $|R\rangle=\frac{1}{\sqrt{2}}\left(
                                    \begin{array}{c}
                                      1 \\
                                      -i \\
                                    \end{array}
                                  \right)$. When the photons pass through the MDM, their
polarization states will be changed by the quarter-wave plates,
which Jones matrix is $P=\left(
                         \begin{array}{cc}
                           1 & 0 \\
                           0 & i \\
                         \end{array}
                       \right)
$, because the prisms used in the MDM act as
a quarter-wave plate on the polarization vector.

Now we will discuss in detail what states the photon will be changed
into under the action of the quarter-wave plates. Table 1 shows the
acting of the quarter-wave plates on the polarization states. From
the table, we can see that the states are always divided into two
sets, diagonal states and circular states. The two diagonal states
$|45^\circ\rangle$ and $|135^\circ\rangle$ can be distinguished by
one measurement, while the two circular states $|L\rangle$ and
$|R\rangle$ can be distinguished by another measurement. But if a
diagonal measurement is performed on a circular photon, the photon
will behave randomly, acting half the time like $|45^\circ\rangle$
and half the time like $|135^\circ\rangle$, and all information
about its circular polarization is lost. Similarly, a random result
is obtained and all information is lost if a circular measurement is
performed on a diagonal photon. Therefore, the traditional BB84
protocol [15] can be used among the users in this scheme.
\begin{table}
  \centering
\caption{Acting of the quarter-wave plates on the four
states,$|45^\circ\rangle$, $|135^\circ\rangle$, $|L\rangle$ and
$|R\rangle$. These states are always divided into two sets, diagonal
states and circular states.}\label{Table 1}
\begin{tabular}{ccccc}

\hline
   & $|45^\circ\rangle$ & $|135^\circ\rangle$ & $|L\rangle$ & $|R\rangle$ \\
\hline
  $P$ & $|L\rangle=\frac{1}{\sqrt{2}}\left(
                                    \begin{array}{c}
                                      1 \\
                                      i \\
                                    \end{array}
                                  \right)$
   & $|R\rangle=\frac{1}{\sqrt{2}}\left(
                                    \begin{array}{c}
                                      1 \\
                                      -i \\
                                    \end{array}
                                  \right)$ & $|135^\circ\rangle=\frac{1}{\sqrt{2}}\left(
                                    \begin{array}{c}
                                      1 \\
                                      -1 \\
                                    \end{array}
                                  \right)$ & $|45^\circ\rangle=\frac{1}{\sqrt{2}}\left(
                                    \begin{array}{c}
                                      1 \\
                                      1 \\
                                    \end{array}
                                  \right)$ \\
  $P$$\cdot$$P$ & $|135^\circ\rangle=\frac{1}{\sqrt{2}}\left(
                                    \begin{array}{c}
                                      1 \\
                                      -1 \\
                                    \end{array}
                                  \right)$ & $|45^\circ\rangle=\frac{1}{\sqrt{2}}\left(
                                    \begin{array}{c}
                                      1 \\
                                      1 \\
                                    \end{array}
                                  \right)$ & $|R\rangle=\frac{1}{\sqrt{2}}\left(
                                    \begin{array}{c}
                                      1 \\
                                      -i \\
                                    \end{array}
                                  \right)$ & $|L\rangle=\frac{1}{\sqrt{2}}\left(
                                    \begin{array}{c}
                                      1 \\
                                      i \\
                                    \end{array}
                                  \right)$ \\
  $P$$\cdot$$P$$\cdot$$P$ & $|R\rangle=\frac{1}{\sqrt{2}}\left(
                                    \begin{array}{c}
                                      1 \\
                                      -i \\
                                    \end{array}
                                  \right)$ & $|L\rangle=\frac{1}{\sqrt{2}}\left(
                                    \begin{array}{c}
                                      1 \\
                                      i \\
                                    \end{array}
                                  \right)$ & $|45^\circ\rangle=\frac{1}{\sqrt{2}}\left(
                                    \begin{array}{c}
                                      1 \\
                                      1 \\
                                    \end{array}
                                  \right)$ & $|135^\circ\rangle=\frac{1}{\sqrt{2}}\left(
                                    \begin{array}{c}
                                      1 \\
                                      -1 \\
                                    \end{array}
                                  \right)$ \\
  $\cdots$ & $\cdots$ & $\cdots$ & $\cdots$ & $\cdots$ \\
  \hline
\end{tabular}

\end{table}

Based on the above OAM-QKD scheme, a quantum network, in which users
can share private information with each other, will be designed
easily. To illustrate the network more clearly, we take the
four-user net for example, as shown in Fig.3(a). Alice, Bob,
Charley, and David are connected by the MDM, whose addresses are
$\ell=4$, $\ell=2$,$\ell=3$ and $\ell=1$ respectively. Any one, who
wants to share information with Alice, for example, ought to prepare
the photons in the $\ell=4$ OAM states. On the other hand, to make
others know the signal photons are sent by Charley, Charley should
sent two light pulses in right-handed, left-handed polarization
respectively before sharing information. The receiver will perform
the corresponding measurement on the fist two photons. It's the
similar communication procedure for other users. As one would
expect, the MDM just helps to connect with two intended users, and
the unconditional security in quantum communications will not
change. Therefore, the usual point-to-point QKD protocols can be
used by the users in the network. Fig.3(b) gives a possible way to
put the photons sent by Alice, Bob, Charley and David into a single
in-port. In Fig.3(b), M1 and M2 are two rotating mirrors controlled
by a program logic control system. When Charley intends to transmit
photons into the in-port, rotation angle of the two mirrors will be
set $\alpha_{1}=\pi/4$ and $\alpha_{2}=-\pi/4$ respectively.
Similarly, the angle ($\alpha_{1}=\pi/4$,$\alpha_{2}=-\pi/2$) is for
David, ($\alpha_{1}=0$,$\alpha_{2}=-\pi/2$) is for Bob, and
($\alpha_{1}=3\pi/4$,$\alpha_{2}=-\pi/2$) is for Alice. More
generally, an intelligent system must be designed to change not only
the rotation angle but also the positions of the two mirrors, since
the incident light is not always from the same direction. We are
ready to do some related work soon.
\begin{figure}
  \includegraphics{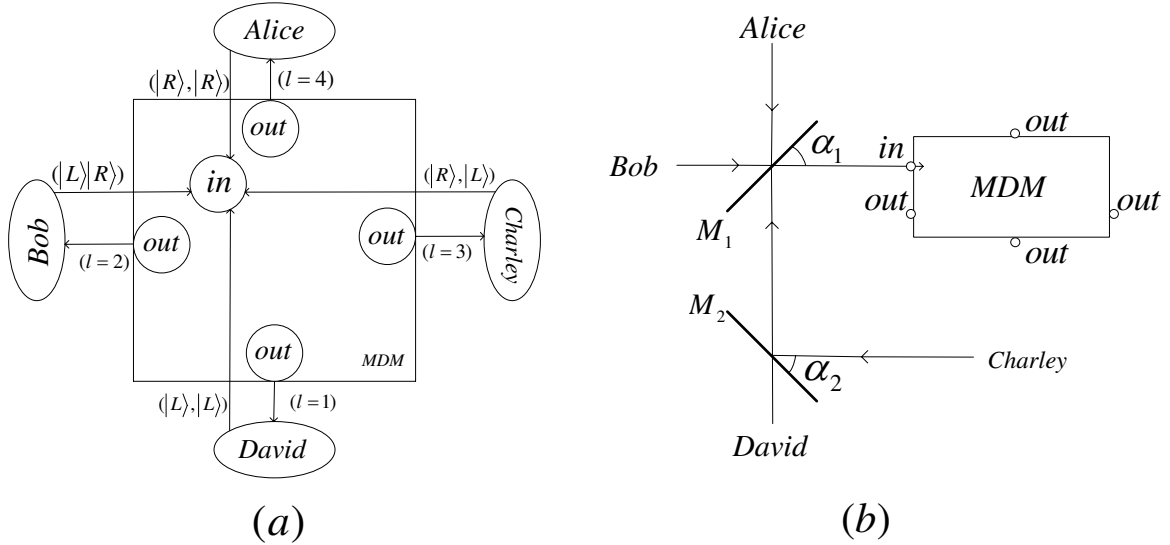}\centering\\
  \caption{(a)The sketch of a four-user network;(b)A possible way to put four separated lights into a single beam line. M1 and M2 are two rotating mirrors.}\label{Fig.3 }
\end{figure}
\label{}
\section{Discussion and Summary}
Though the optical fiber is the main transmission medium for optical
communication, it still requires to perform QKD through free space,
since the satellite technology has been popularized nowadays.
Besides, it is difficult to distribute the quantum key through more
than 200 km with fiber due to the losses and the low single-photon
detection efficiency. And free-space QKD is now attracting
increasing attention[16-18] because of its low-transmission loss. It
is a good idea to implement our network scheme among the relay
stations like the moving satellites. For one thing, the OAM states
are invariant under rotations about the propagation direction,
making this implementation independent of the alignment between the
reference frames of the sender and receiver, and hence appealing for
free space QKD[19]. For another, the essentially nonbirefringe
nature of the atmosphere at optical wavelengths allows the faithful
transmission of the single photon polarization states used in our
network scheme. Therefore, our scheme is worthy to be studied
further.

However, there are challenges to implement our scheme in free space
QKD. First, the atmospheric turbulence will destroy the vortex
structure of LG modes even for weak turbulence[20]. Second, before
applying our newly-proposed scheme to a large scale QKD network, we
should find out a practical way to put more separated lights into a
single beam line. Third, the problems of general free space QKD,
such as beam wander and background photon counts, will present to
our scheme also. The three difficulties mentioned above will become
quite meaningful subjects in scientific research about free space
QKD. As to the first problem, the distortions created by atmospheric
turbulence on OAM photon states[20] could be corrected using
two-dimensinal filtering techniques that have been proposed for
image recovering under various degradation mechanisms[21]. Regarding
the second question, we are pleased to see that it opens up a whole
new topic for all scientists, and we are preparing to do some
related research next. As for the third point, researchers have put
forward a series of possible ways, including narrow filters[22][23],
spatial filtering[24], and adaptive optics[25], to render the
transmission and detection problems over these years. With the above
interesting problems, the pace of scientific research will not stop.

In conclusion, we have devised a novel network scheme based on OAM
and SAM of photons, which is particularly appealing for the free
space QKD. We also describe a four-user experimental scheme of this
efficient quantum network in detail. Users in this network can
freely exchange information with each other.

\label{}

\section{Acknowledgements}
This work is supported by the State Key Development Program for
Basic Research of China (Grant No.2007CB307001).

\label{}

\end{document}